\documentclass[preprint,showpacs,aps,amsfonts,amsmath,amssymb,superscriptaddress,altaffilletter]{revtex4}
\usepackage{graphicx}

\begin{document}
\author{Yang Huang}
\author{Jun She}
\affiliation{Department of Physics, Peking University, Beijing
100871, China}

\author{Bo-Qiang Ma}
\email{mabq@phy.pku.edu.cn} \affiliation{Department of Physics and
MOE Key Laboratory of Heavy Ion Physics, Peking University, Beijing
100871, China}

\title{Single Target-Spin Asymmetry in Semi-Inclusive Deep Inelastic
  Scattering on Transversely Polarized Nucleon Target }

\begin{abstract}

We use a new set of Collins functions to update a previous
prediction on the azimuthal asymmetries of pion productions in
semi-inclusive deep inelastic scattering~(SIDIS) process on a
transversely polarized nucleon target. We find that the calculated
results can give a good explanation to the HERMES experiment with
the new parametrization, and this can enrich our knowledge of the
fragmentation process. Furthermore, with two different approaches of
distribution and fragmentation functions, we present a prediction on
the azimuthal asymmetries of pion and kaon productions at the
kinematics region of the experiments E06010 and E06011 planned at
Jefferson Lab~(JLab). It is shown that the results are insensitive
to the models for the pion case. However, the results for kaon
production are sensitive to different approaches of distribution and
fragmentation functions. This is helpful to clarify some points in
the study of the azimuthal spin asymmetries and fragmentation
functions in hadronization processes.
\end{abstract}


\maketitle

\section{Introduction}
The history of single spin asymmetries~(SSA) can date back to the
1970s when significant SSA were observed in
$pp\rightarrow\Lambda^{\uparrow} X$ ~\cite{Bunce}. In the early
1990s, large asymmetries in $p^{\uparrow}p \rightarrow\pi X$ were
found at FNAL~\cite{fnal}. However, there were no satisfactory
theory to describe the phenomena, and pQCD theory had nothing to do
then. In recent years, SSA phenomenon were also observed in
semi-inclusive deep inelastic scattering~(SIDIS) processes, which
had attracted many interests, particularly in the case where the
transverse polarized targets are used. For example, HERMES
collaboration~\cite{HERMES,HERMES1} has reported the observation of
the azimuthal asymmetries in single-pion productions on both
longitudinally and transversely polarized hydrogen targets. More
recently, COMPASS collaboration~\cite{COMPASS} has published their
results off a transversely polarized deuterium targets as a
complementary measurement to HERMES experiment. On the theoretical
side, by taking into the account of the transverse parton momenta
inside the nucleon~\cite{Co So,Ji Ma Yuan,Co Metz}, these
asymmetries are now assumed to have correlations with the concept of
transversity~\cite{bdr} which we are not familiar with so far.

The idea~(though not the term) of transversity was put forward by
Ralston and Soper via the Drell-Yan process~\cite{Soper}, where they
introduced the concept of parton transverse polarization. A
clarification of transversity on the role of chiral-odd parton
distributions and the general twist was provided by Jaffe and Ji in
Ref.~\cite{Ja Ji}. Detailed twist clarification for various parton
distributions can be found in Refs.~\cite{bdr,rat}. Now we know that
at the leading twist~(twist 2), three fundamental quark
distributions provide a complete description of quark momentum and
spin in the nucleon. In the last forty years, two of them, the
unpolarized and longitudinal polarized parton distributions($q_i$
and $\Delta q_i$) have been precisely measured, yet the third type,
the ``transversity distribution ($\delta q_i$)'', is still little
known both theoretically and experimentally. The difficulty in
experiments lies in its chiral-odd property. In the helicity basis,
$\delta q_i$ represents a quark helicity flip, which cannot occur in
any hard process for massless quarks within QED or QCD~\cite{XDJ}.
This chiral-odd property makes it inaccessible in inclusive deep
inelastic scattering (DIS). However, several researches have shown
that the transversity distribution can manifest itself in
semi-inclusive deep inelastic scattering (SIDIS)
reactions~\cite{jc:1993,Jaffe prl:1998,Kotzinain:1995}, where the
transversity distribution function couple with a chiral-odd
fragmentation function---the Collins function~\cite{jc:1993}. This
was verified by observing the SSA in HERMES and COMPASS experiments
recently. Results from both HERMES and COMPASS have offered us a
first glimpse at transversity distributions on u and d quarks in
proton.

If studying SSA more explicitly, we would find that these
asymmetries can be explained ~\cite{Boer Mulder} in terms of both
the Sivers~\cite{DW:1991,sj:2002,jc:2002,xj:2003} and
Collins~\cite{jc:1993} effects. The Sivers effect involves the so
called Sivers function and the ordinary fragmentation function,
while the Collins effect involves the transversity distribution
function and the Collins fragmentation function. The two competing
contributions have different kinematic dependence. The Collins
effect depends on \textit{y} and strongly correlated with
\textit{z}, but the Sivers effect is independent of \textit{y} and
not strongly correlated with \textit{z}. Also, they have different
azimuthal angle dependence: the Sivers effect is proportional to
$\sin(\phi-\phi_s)$, while the Collins effect is proportional to
$\sin(\phi+\phi_s)$, where the definition of $\phi$ and $\phi_s$ can
be found in Ref.\cite{Boer Mulder}. Thus we can distinguish the two
effects in experiment without much difficulty. In our paper, we are
interested in the transversity information provided by SSA, so we
only concentrate on the Collins effect in this paper. In the Collins
effect,the transversity distribution
 function describes the correlation
between the spin of quarks and spin of nucleons, and the so called Collins function
describes the fragmentation of transversely polarized quarks into
unpolarized hadrons. Both functions are important in our numerical calculations and
will be discussed in the next section.

With this understanding, a lot of
studies~\cite{am:2000,av:2000,as:2000,db:2000,mb:2000,ma:2000,ks:2000,
bq:2001,ka:1998,ed:2000,va:2001,av:2001,bq:,bq:2} have attempted to
explain the HERMES measurement or to give more predictions.
Particularly, the article~\cite{bq:} predicted the Collins azimuthal
asymmetries (Fig.~\ref{fig1}) on a transversely polarized hydrogen
target in HERMES kinematics, yet the prediction is not consistent
with the new released HERMES results~\cite{HERMES1}. However, as the
authors pointed out in the paper that the ``unfavored'' process
might lead to a sizable effect, which is in conflict with our
general understanding and might result in the inconsistency.
Recently, the authors of Ref.~\cite{fe:2005} proposed that the
polarized ``unfavored'' Collins function is approximately equal to
the ``favored'' one, but with an opposite sign. They fitted the
HERMES data~\cite{HERMES1} and obtained new sets of parametrization.
We will shown in this paper that the updated results of
Ref.~\cite{bq:} with these new parametrization of the Collins
function, are also consistent with the HERMES data (Fig.~\ref{fig2})
. Both calculations imply that the ``unfavored'' Collins function
may play an important role in the fragmentation processes.

At Jefferson Lab~(JLab), experiments E06010 and E06011 will perform
the measurement of single spin asymmetries with a transversely
polarized $^3$He target which is an effectively transversely
polarized neutron target, so it may provide a direct measurement of
the neutron transversity distributions. In our paper,
 we will present the predictions of the azimuthal
asymmetries in the JLab kinematics region. The influence due to the
``unfavored'' Collins function is included in the prediction. We
conclude that the comparison of the data with the
prediction will be able to provide constraints on both the
Collins functions and transversity distributions, but not a
pure measurement of the transversity distributions.

\section{Quark Distribution and Fragmentation functions}

At the leading twist, the differential cross section for a SIDIS
reaction includes the contributions from the unpolarized part, the
Collins effect and the Sivers effect. It can be written as
\begin{eqnarray}
d\sigma=d\sigma_{UU}-d\sigma^{\mathrm{Collins}}_
{UT}-d\sigma^{\mathrm{Sivers}}_{UT}, \label{duu}
\end{eqnarray}
where the beam is not polarized and the target is transversely
polarized.

Under the factorization assumption, the cross sections in
Eq.~(\ref{duu}) can be expressed as product of parton distribution
and fragmentation functions~\cite{am:1997}. Here we only give the
unpolarized and the Collins terms as follows,
\begin{eqnarray}
\frac{d\sigma^{\ell+N\rightarrow\ell^\prime+h+X}}{dx dy dz
d\phi^\ell d^2p^h_\perp}&=&\frac{4\pi
\alpha^2sx}{Q^4}(1-y+\frac{y^2}{2})\sum_{q}e^2_q  q(x)
D^q_1(z,P^2_{h\perp})\nonumber\\
&-&|S_T|(1-y)\sin(\phi_h^\ell+\phi_s^\ell)\sum_{q}e^2_q \delta
q(x)H_1^{\perp(1) q}(z,P^2_{h\perp}).\label{cross section}
\end{eqnarray}
Here $S_T$ is the target transverse polarization; $e_q$ is the
charge of the quark with flavor $q$. $\phi^\ell_h$ is the angle
between the hadron lane and the lepton plane. $\phi^\ell_s$ is the
angle between the $\vec{S}$ and the lepton plane. $q(x)$ and $\delta
q(x)$ represent the momentum and transversity distribution functions
of the nucleon target respectively. By comparing the cross sections
with the opposite polarized target, one obtains the single spin
asymmetry
\begin{eqnarray}
A_{UT}\equiv\frac{d\sigma(\vec{S}_T)-d\sigma(-\vec{S}_T)}{d\sigma(\vec{S}_T)+d\sigma(-\vec{S}_T)}=
A_T^{Collins}(x,y,z,p^h_\perp)\cdot \sin(\phi^\ell_h+\phi^\ell_s).
\end{eqnarray}
After integrating the intrinsic transverse momentum $P_{h\perp}$ and
the kinematical variable $y$ and $z$, we can get the formula for
calculating the $x$ dependence of the Collins asymmetry~(denoted as
$A_T(x)$)
\begin{eqnarray}
A_T(x)=-|S_T|\frac{\sum_q e^2_q \int dz dy (1-y)\delta
q(x)H_1^{\perp(1)q}(z) }{\sum_q e^2_q \int dz dy(1-y+y^2/2) q(x)
D_1^q(z)}. \label{a1}
\end{eqnarray}
 $D^q_1(z)$ is the fragmentation
function for an unpolarized quark with flavor $a$ into a hadron,
defined as
\begin{eqnarray}
D_1^q(z)\equiv \int d^2P_{h\perp} D_1^q(z,P_{h\perp}^2).
\end{eqnarray}
$H_1^{\perp(1) q}(z)$ is the so called Collins function, defined as
\begin{eqnarray}
H_1^{\perp(1) q}(z)\equiv \int d^2
P_{h\perp}\frac{-P^2_{h\perp}}{z^2M^2_h}H_1^{\perp
q}(z,P^2_{h\perp}).
\end{eqnarray}
The kinematical variable $y$ can be calculated from
\begin{eqnarray}
Q^2=sxy,
\end{eqnarray}
with $s$=51.8 ~GeV$^2$.

To get the transversity distribution function, we use two models in
this paper, the quark-diquark model~\cite{rp:1972,fe:1973,bq:1996}
and a pQCD based counting rule
analysis~\cite{gr:1975,rb:1974,sj:1995,mg:2001,mh:2004}. In the
quark diquark model case, the transversity distributions are given
by
\begin{eqnarray}
\delta
u_v(x)&=&[u_v(x)-\frac{1}{2}d_v(x)]W_s(x)-\frac{1}{6}d_v(x)W_v(x);\nonumber\\
\delta d_v(x)&=&-\frac{1}{3}d_v(x)W_v(x).
\end{eqnarray}
$W_s(x)$ and $W_v(x)$ are the Melosh-Wigner rotation factors for
spectator scalar and vector diquarks, which come from the
relativistic effect of quark transversal
motions~\cite{ma1991,Sch97}. For the pQCD based analysis, we adopt
the parametrization
\begin{eqnarray}
u_v^{pQCD(x)}=u_v^{para}(x),~~~~~~d_v^{pQCD}(x)=\frac{d_v^{th}(x)}{u_v^{th}(x)}u_v^{para}(x),\\
\delta u_v^{pQCD}(x)=\frac{\delta
u_v^{th}(x)}{u_v^{th}(x)}u_v^{para}(x),~~~~~~\delta
d_v^{pQCD}(x)=\frac{\delta d_v^{th}(x)}{u_v^{th}(x)}u_v^{para}(x),
\end{eqnarray}
where the superscripts ``th'' means the theoretical calculation in
the pQCD analysis~\cite{bq:2001,bq:}, and ``para''means the input
from parametrization. The CTEQ~\cite{hl:2000} parametrization is
served as the input for both models to get the unpolarized parton
distribution functions. Detailed constructions of the quark
distributions can be found in Refs.~\cite{bq:,bq:2001,bq:2000}.

Both of the two models have successfully predicted the longitudinal
polarized parton distribution functions~\cite{xc}. But two models
give different distribution functions at $x\rightarrow{1}$: the pQCD
based counting rule analysis~\cite{sj:1995} predicts
$\delta{d(x)}/d(x)\rightarrow {1}$, while the SU(6)
quark-spectator-diquark model~\cite{bq:1996} predicts
$\delta{d(x)}/d(x)\rightarrow{-1/3}$. In a recent
literature~\cite{Anselmino}, the author extracted the transversity
distribution for $u$ and $d$ quarks from the now available data.
They found that $\delta u(x)$ and $\delta d(x)$ turned out to be
opposite in sign, with $|\delta d(x)|$ smaller than $|\delta u(x)|$.
This seems to be coincidence qualitatively with the SU(6)
quark-diquark model. But there it predicts $\delta
d(x)/d(x)\rightarrow 0$ when $x\rightarrow 1$, which is coincidence
quantitatively with neither models we used in our paper. The
correctness of different parametrization is still unclear, and need
to be checked by more experiments.

As to the ordinary fragmentation functions for $\pi^{\pm}$, we
follow the parametrization of $D(z)$ given by Kretzer, Leader and
Christova~\cite{sk:2001}:
\begin{eqnarray}
D^{\pi^{\pm}}(z)=0.689z^{-1.039}(1-z)^{1.241},\nonumber\\
\hat{D}^{\pi^{\pm}}(z)=0.217z^{-1.805}(1-z)^{2.037}, \label{d4}
\end{eqnarray}
and for $K^{\pm}$, we have~\cite{Binnewies}
\begin{eqnarray}
D^{K^{\pm}}(z)=0.31z^{-0.98}(1-z)^{0.97},\nonumber\\
\hat{D}^{K^{\pm}}(z)=1.08z^{-0.82}(1-z)^{2.55},
\end{eqnarray}
 where $D(z)$
denotes the ``favored'' fragmentation function and $\hat D(z)$
denotes the ``unfavored'' one.

The Collins fragmentation function $H_1^{\perp(1)}(z)$, which
describes the transition of a transversely polarized quark into a
pion, is theoretically little known and has not yet been measured.
In Ref.~\cite{jc:2002}, Collins suggested a parametrization:
\begin{eqnarray}
A_C(z,k_{\perp})=\frac{|k_\perp|{H}_1^{{\perp}q}(z,z^2k^2_\perp)}
{M_hD^q_1(z,z^2k^2_\perp)}=\frac{M_c|k_\perp|}{M^2_C+|k^2_\perp|},
\label{h}
\end{eqnarray}
where $M_c=0.3\sim1$ GeV. In HERMES analysis, $M_c=0.7$ GeV is taken
as a rough estimate~\cite{va:2001}. The transverse momentum
dependence of $D^q_1(z,z^2k^2_{\perp})$ is assumed to have a
Gaussian type shape:
\begin{eqnarray}
D^q_1(z,z^2k^2_{\perp})=D^q_1(z)\frac{R^2}{\pi{z}^2}\exp(-R^2k^2_\perp).
\label{dd1}
\end{eqnarray}
One can obtain
\begin{eqnarray}
H^{\perp(1)q}_1(z)=D^q_1(z)\frac{M_c}{2M_h}(1-M^2_CR^2\int^\infty_0dx\frac{\exp(-x)}{x+M^2_CR^2}),
\label{col}
\end{eqnarray}
with $R^2=z^2/\langle P^2_{h\perp}\rangle$, and $\langle
P^2_{h\perp}\rangle=z^2\langle k^2_\perp\rangle$ being the mean
square momentum that the detected hadron acquires in the quark
fragmentation process with $\langle P^2_{h\perp}\rangle=0.25$
$\mathrm{GeV}^2$ according to HERMES~\cite{va:2001}.

Recently, Vogelsang and Yuan suggested another model~\cite{fe:2005}
which assumes a stronger constraint for the pion Collins functions:
\begin{eqnarray}
H_1^{\perp{(1)}\pi^+}(z)+H_1^{\perp{(1)}\pi^-}(z)+H_1^{\perp{(1)}\pi^0}(z)\approx{0}.
\label{h1}
\end{eqnarray}
If we consider the isospin and charge symmetry relations between the
different fragmentation functions, we can have the following
relations,
\begin{eqnarray}
H_u^{\perp{(1)}\pi^+}(z)&=&H_d^{\perp{(1)}\pi^-}(z)=H_{\bar{u}}^{\perp{(1)}\pi^-}(z)=
H_{\bar{d}}^{\perp{(1)}\pi^+}(z)=H(z),\nonumber\\
H_u^{\perp{(1)}\pi^-}(z)&=&H_d^{\perp{(1)}\pi^+}(z)=H_{\bar{u}}^{\perp{(1)}\pi^+}(z)=
H_{\bar{d}}^{\perp{(1)}\pi^-}(z)=\hat{H}(z),\nonumber\\
H_u^{\perp{(1)}\pi^0}(z)&=&H_d^{\perp{(1)}\pi^0}(z)=H_{\bar{u}}^{\perp{(1)}\pi^0}(z)=
H_{\bar{d}}^{\perp{(1)}\pi^0}(z)=\frac{1}{2}[H(z)+\hat{H}(z)].
\label{hu}
\end{eqnarray}
Substituting (\ref{hu}) into (\ref{h1}), one obtains
\begin{eqnarray}
H(z)+\hat{H}(z)\approx0, \label{hh}
\end{eqnarray}
which means that the ``unfavored'' Collins function is approximately
equal to the ``favored'' one, but with an opposite sign. This is in
conflict with almost all the theoretical analysis before, and it
still needs to be further checked by experiment. In
Ref.~\cite{fe:2005}, the parametrization of Collins functions is
advanced in the forms,
\begin{eqnarray}
\textrm{Set I:}~H(z)=C_fz(1-z)D(z),~\hat{H}(z)=C_uz(1-z)D(z),\label{setI}\nonumber\\
\textrm{Set II:}~H(z)=C_fz(1-z)D(z),~\hat{H}(z)=C_uz(1-z)\hat{D}(z),\label{setII}
\end{eqnarray}
 The authors of Ref.~\cite{fe:2005} fitted
the parametrization to the HERMES data~\cite{HERMES} and obtained
\begin{eqnarray}
\mathrm{Set~I}&:& C_f=-0.29\pm0.04,~C_u=0.33\pm0.04,
\nonumber\\\mathrm{Set~II}&:& C_f=-0.29\pm0.02,~C_u=0.56\pm0.07.
\label{set}
\end{eqnarray}

The difference between these two sets of parametrization is that in
Set I, both favored and unfavored Collins functions are
parameterized in terms of favored unpolarized quark fragmentation
function, while in Set II the unfavored Collins function is
parameterized in terms of the unfavored unpolarized quark
fragmentation function. Also in the literature~\cite{Anselmino}, the
authors provide their own parametrization to the Collins functions,
with more parameters compared to the parametrization in
Ref.~\cite{fe:2005}. As the author pointed that their
parametrization is agree with the extractions obtained in
Ref.~\cite{fe:2005}, we will use the parametrization with less
parameters provided in Ref.~\cite{fe:2005} in our paper.

This parametrization was obtained from the pion data in
Ref.~\cite{fe:2005}. In our paper, we assume that this
parametrization is also right for the kaon case. In other words, we
assume that with this parametrization, the information depending on
the final hadron states is only contained in the ordinary
fragmentation functions.

\section{Comparison with the HERMES data}

\begin{figure}
\center
\includegraphics[width=8cm]{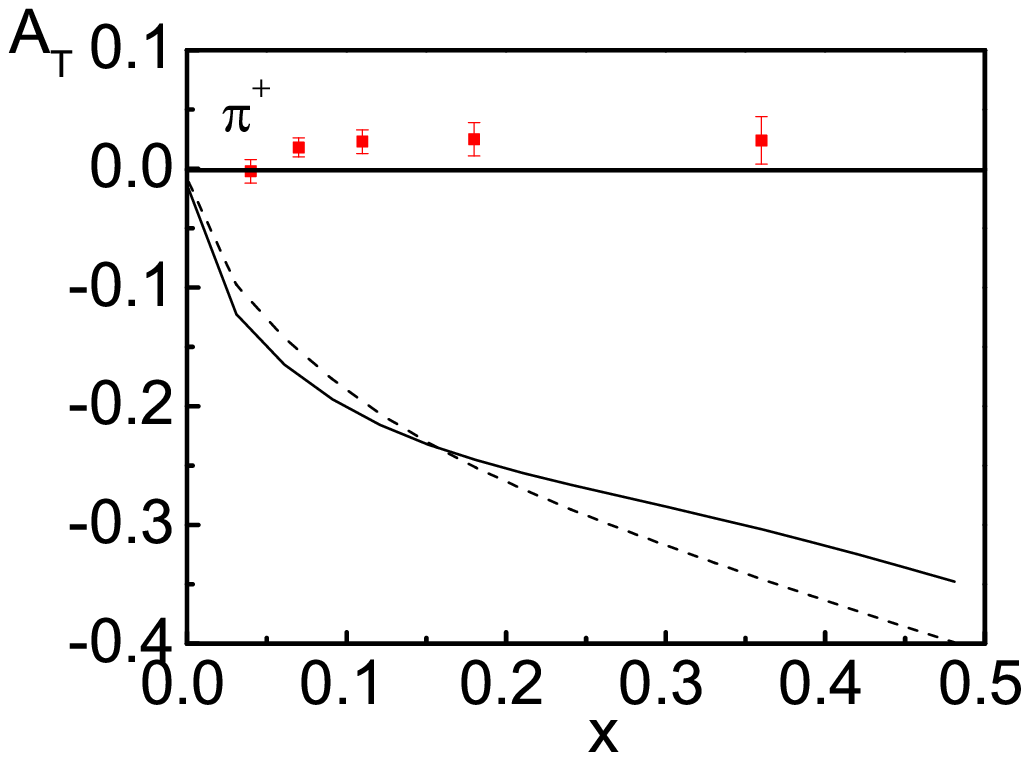}
\includegraphics[width=8cm]{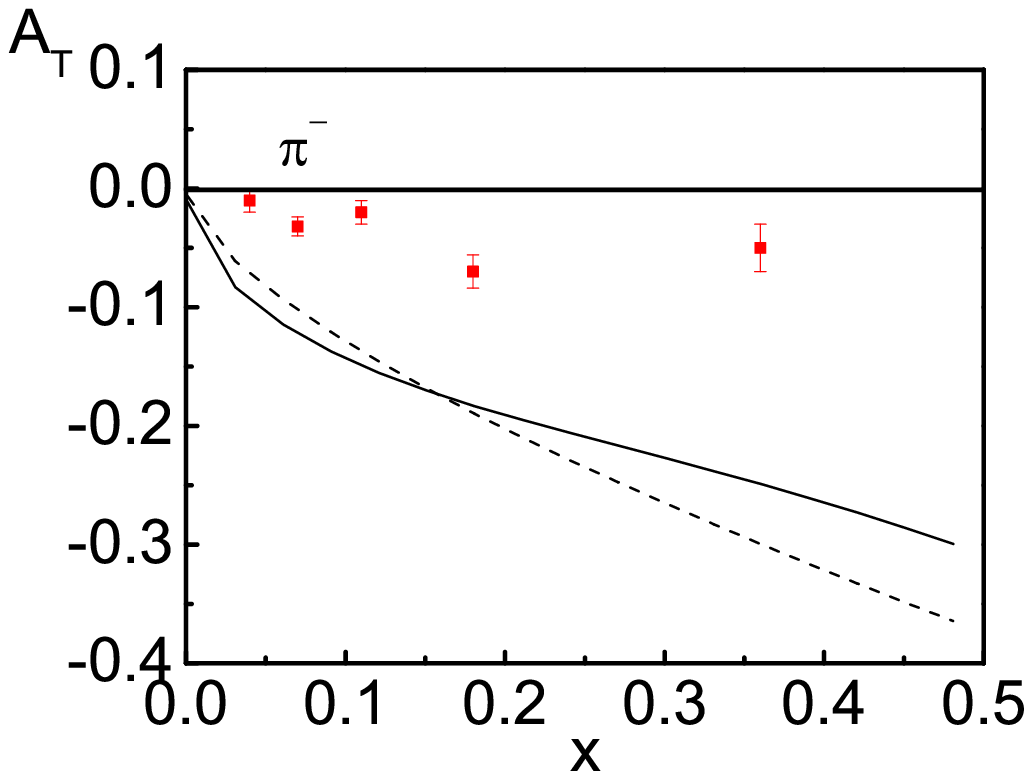}
\caption{The azimuthal asymmetries $A^{\sin(\phi+\phi_s)}_{T}$ for
the semi-inclusive $\pi^+$ and $\pi^-$ productions in deep inelastic
scattering of unpolarized charged lepton on a transversity polarized
$proton$ target in the HERMES kinematics region, with the the
parametrization for the Collins function of (\ref{col}). The solid
and dashed curves correspond to the calculated results for
quark-diquark model and the pQCD based analysis, respectively.}
\label{fig1}
\end{figure}

\begin{figure}
\center
\includegraphics[width=8cm]{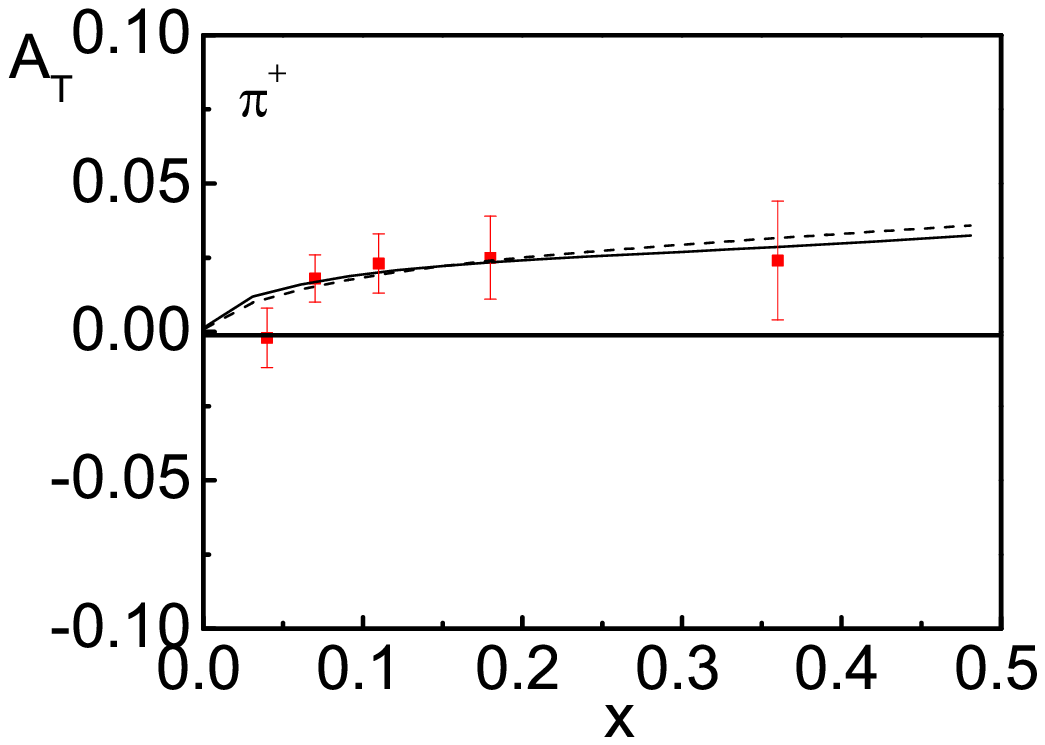}
\includegraphics[width=8cm]{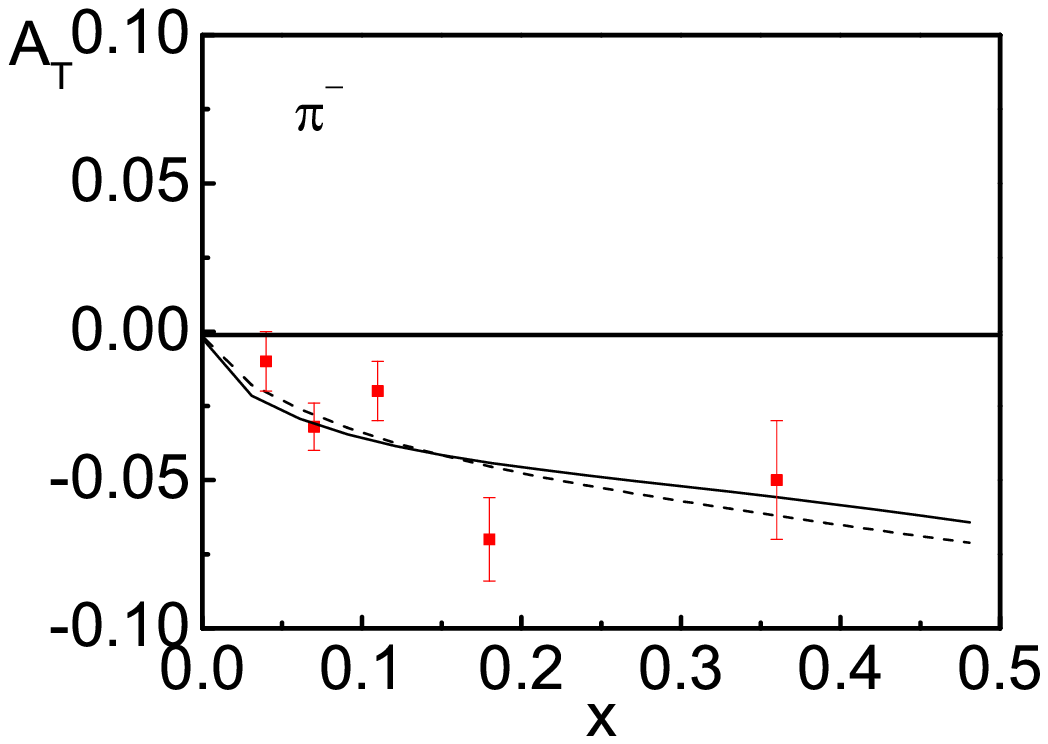}
\includegraphics[width=8cm]{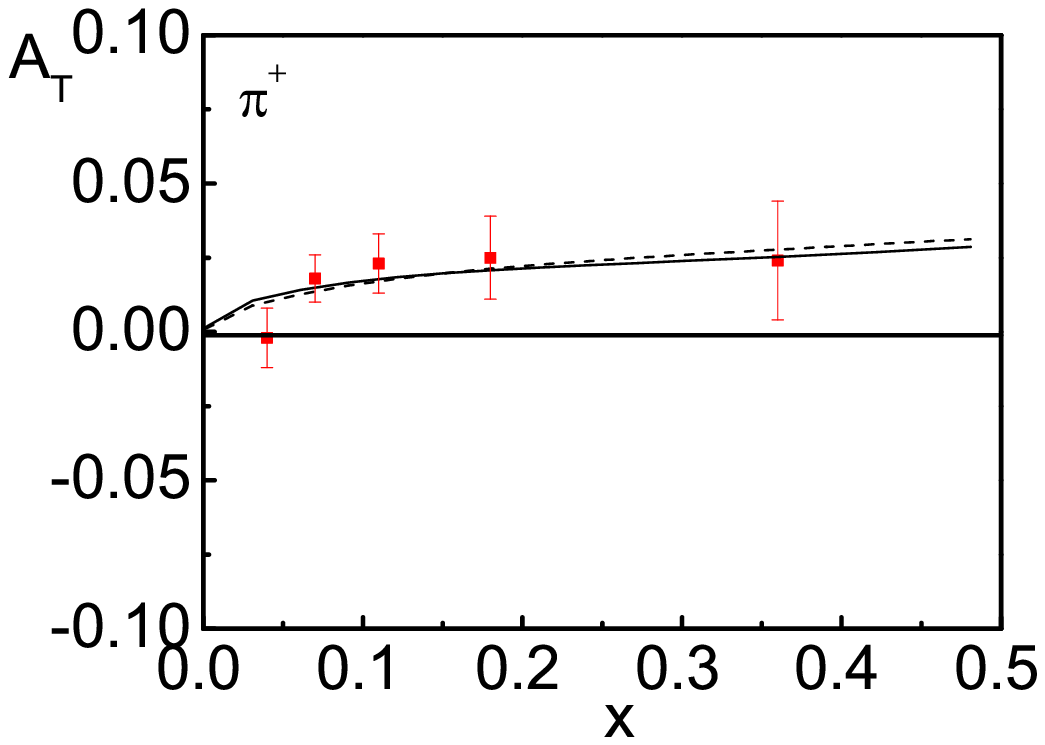}
\includegraphics[width=8cm]{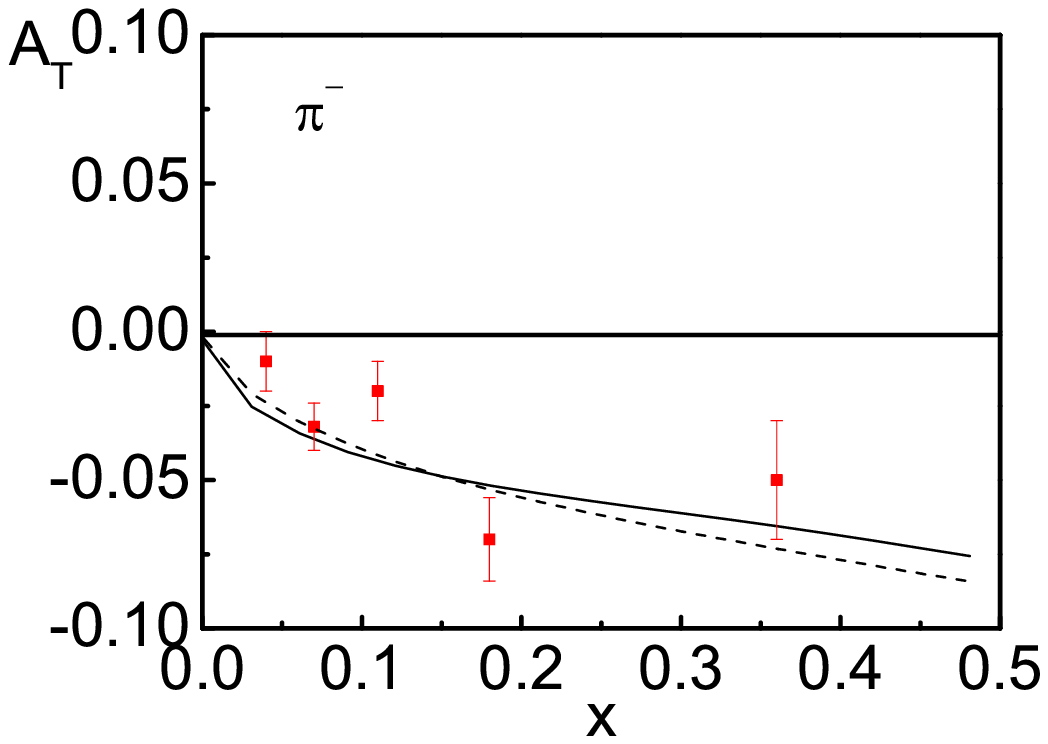}
\caption{The azimuthal asymmetries $A^{\sin(\phi+\phi_s)}_{T}$ for
the semi-inclusive $\pi^+$ and $\pi^-$ productions in deep inelastic
scattering of unpolarized charged lepton on a transversely polarized
$proton$ target in the HERMES kinematics region. We use the
parametrization of equations (\ref{setI}). The upper row corresponds
to the Set I parametrization while the lower row corresponds to the
Set II parametrization. The solid and dashed curves correspond to
the calculated results for quark-diquark model and the pQCD based
analysis, respectively.} \label{fig2}
\end{figure}

Figs.~{\ref{fig1}} and {\ref{fig2}} show our calculated results on
HERMES measurement as function of $x$, using different sets of
parametrization. First we find that the quark-diquark model and the
pQCD based counting rule analysis give almost the same result in the
intermediate region of \textit{x}. Fig.~\ref{fig1} presents the
results when we use the parametrization as Eq.~(\ref{col}) shows.
However, the calculated results are overestimated compared to the
experimental data. Fig.~\ref{fig2} presents the results when we use
two sets of parametrization of Collins functions given by
Ref.~\cite{fe:2005}, which shows that our calculations
 fit the HERMES data well. As $u$ quark dominates in the proton, so
 $u$ to $\pi^+$ is favored while $u$ to $\pi^-$ is unfavored.
 The calculation
 indicates that both the ``favored'' and ``unfavored'' processes have
 the sizable effect. This is the direct consequence of Eq. \ref{set} where the
favored and unfavored Collins
functions have the approximately equal size but with the opposite signs.
This is quite different
 from the ordinary fragmentation functions where the favored process
 plays much more important roles than the unfavored process. If we ignore
the opposite sign, the numerator of formula (\ref{a1}) is nearly of
the equal size for both the $\pi^+$ and $\pi^-$ productions, but the denominator
is larger in the $\pi^+$ production case than that in the $\pi^-$ production case.
So we expect a larger asymmetry in the $\pi^-$ production and this can
also be seen from Fig.\ref{fig2}. Besides these, we also notice that both the
two sets of parametrization shown in Eq. (\ref{set}) give similar predictions,
so more precise experiments are needed to give the constraints.

\section{Predictions on the JLab experiments}

The JLab experiments E06010 and E06011 will measure the target
single spin asymmetries in the semi-inclusive deep-inelastic
reaction off a transversely polarized $^3$He target for both
$\pi^{\pm}$ and $K^{\pm}$. Duo to the special structure of $^3$He,
which can be considered as a nearly free polarized neutron, JLab
will provide a measure of the neutron structure. Applying the same
technique, we make predictions of azimuthal asymmetries $A_T$ for
$\pi^{\pm}$ and $K^{\pm}$ productions at JLab kinematics.

At JLab, the kinematics region is:
\begin{eqnarray}
2.33 < W < 3.05~\mathrm{GeV}, ~0.19 < x < 0.34,~1.77 < Q^2 <
2.73~\mathrm{GeV^2},~0.37< z < 0.56.~\label{jlab}
\end{eqnarray}
The azimuthal asymmetry of $^3$He target can be expressed by the
neutron and proton asymmetries:
\begin{eqnarray}
A_T(^3He)=P_{^3He}\cdot(f_n\cdot{\eta_n}\cdot{A_T(n)+2f_p\cdot{\eta_p}\cdot{A_T(p)}}),\label{He}
\end{eqnarray}
where $f_n$ and $f_p$ are the effective polarizations of the proton
and the neutron within the $^3$He nucleus, and $P_{^3He}$ is the
polarization of the $^3$He target which is assumed to be $42\%$ in
the experiment. A three-body Fadeev calculation~\cite{cc:1993} shows
that in inelastic scattering reactions, $f_n=0.86\pm0.02$ and
$f_p=-0.028\pm0.004$. $\eta_n$($\eta_p$) in the above formula
represents the ratio of $(e,e'\pi)X$ events on neutron (proton) over
the total $^3$He$(e,e'\pi)X$ events. At JLab, one expects:
$\eta_n\thickapprox0.32$ and $\eta_p\thickapprox0.34$. For the case
of kaon production, we adopt the same value as a rough estimate.

\begin{figure}
\center
\includegraphics[width=8cm]{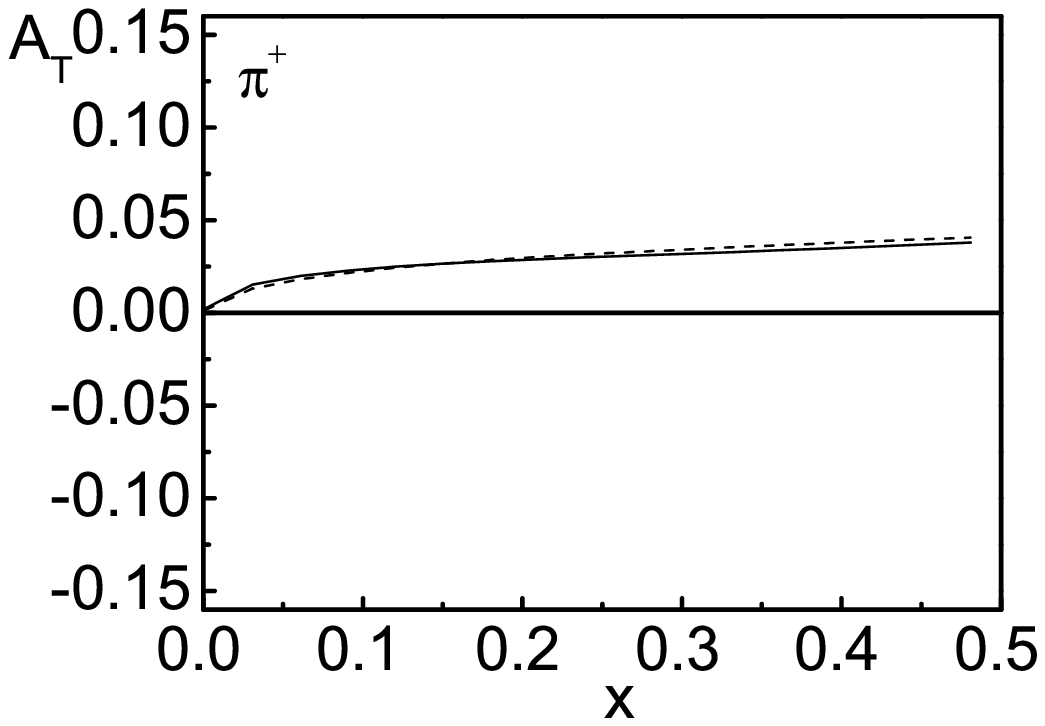}
\includegraphics[width=8cm]{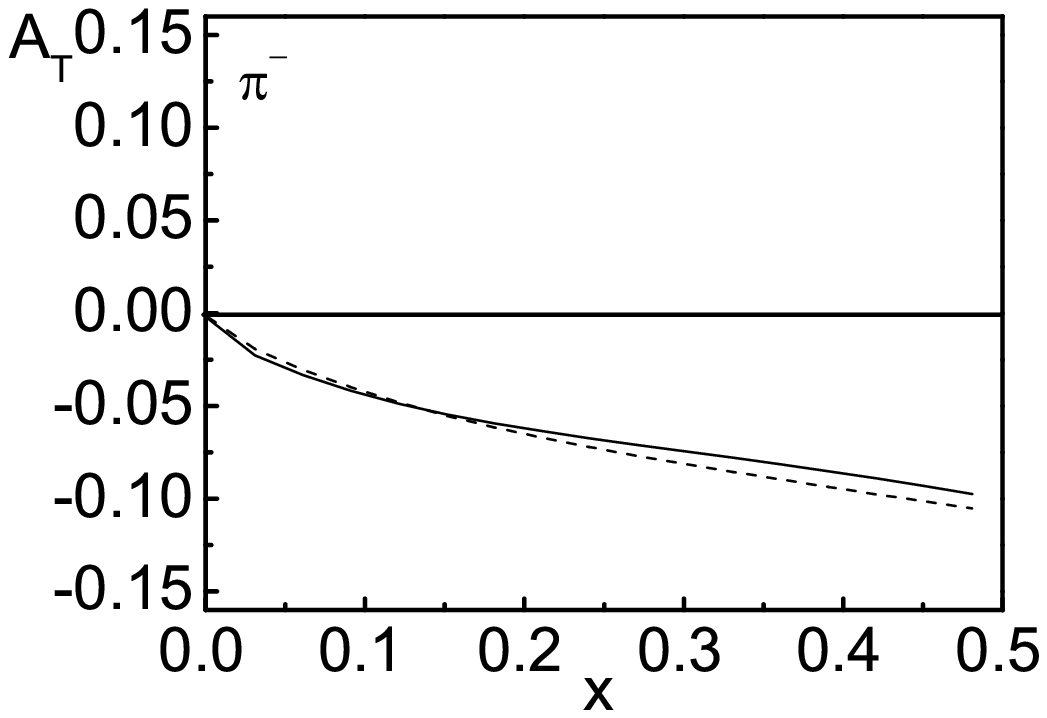}\nonumber\\
\includegraphics[width=8cm]{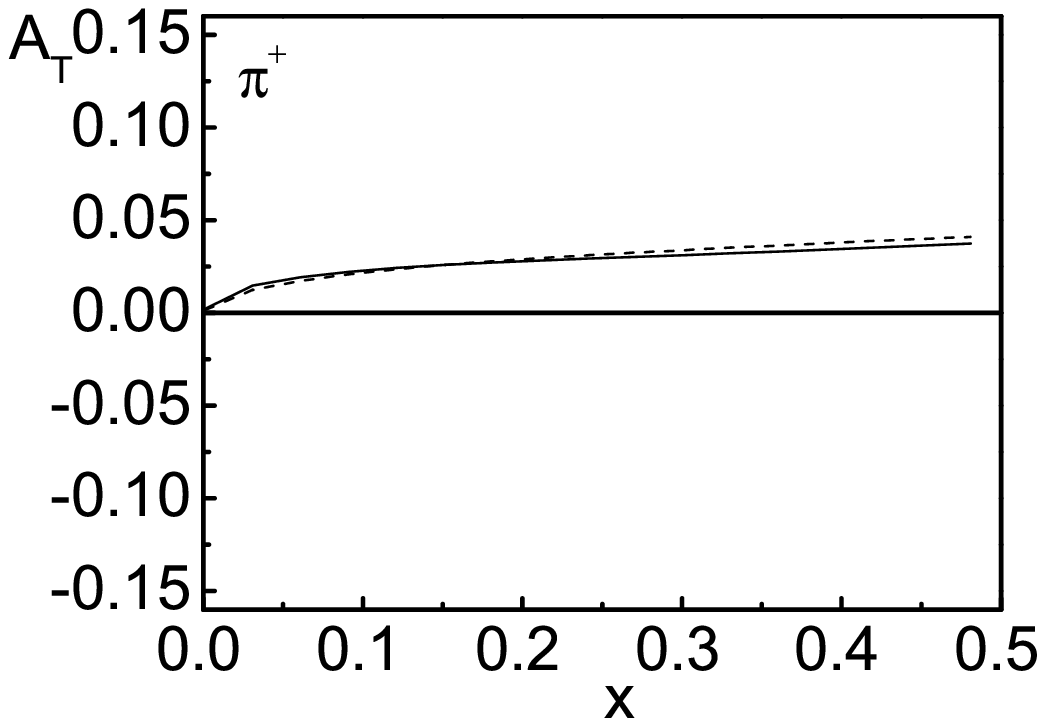}
\includegraphics[width=8cm]{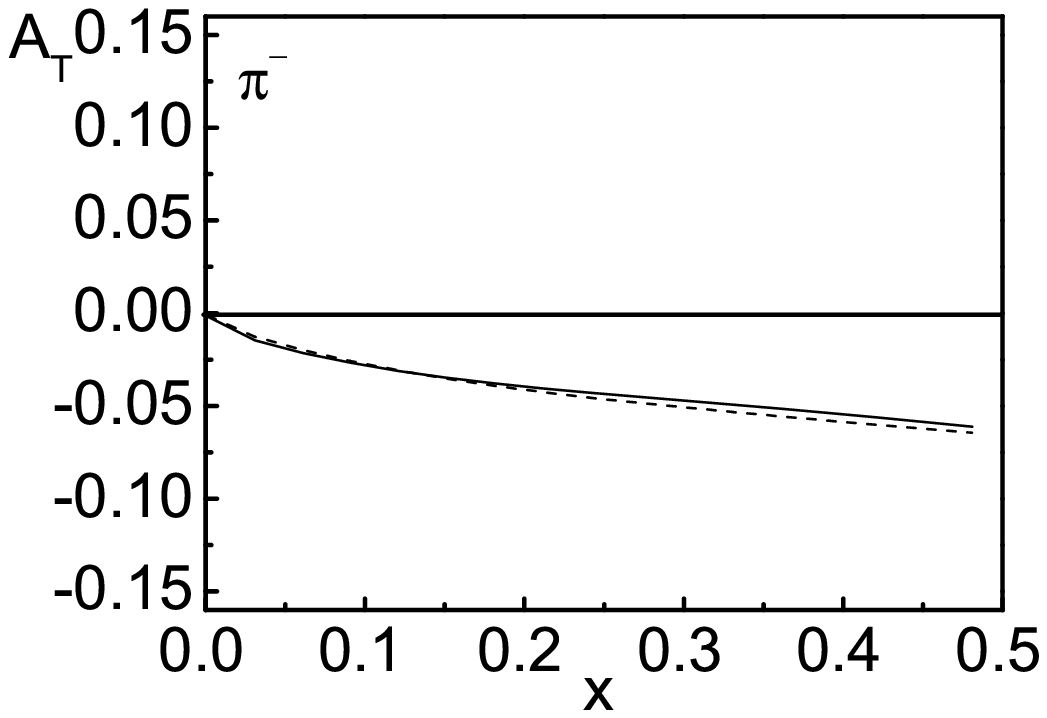}

\caption{The azimuthal asymmetries $A^{\sin(\phi+\phi_s)}_{T}$ for
the semi-inclusive $\pi^+$ and $\pi^-$ productions in deep inelastic
scattering of unpolarized charged lepton on a transversely polarized
$proton$ target with the JLab kinematics cut. We use the
parametrization of equations (\ref{setI}). The upper row corresponds
to the Set I parametrization while the lower row corresponds to the
Set II parametrization. The solid and dashed curves correspond to
the calculated results for quark-diquark model and the pQCD based
analysis, respectively.}
 \label{fig3}
\end{figure}

\begin{figure}
\center
\includegraphics[width=8cm]{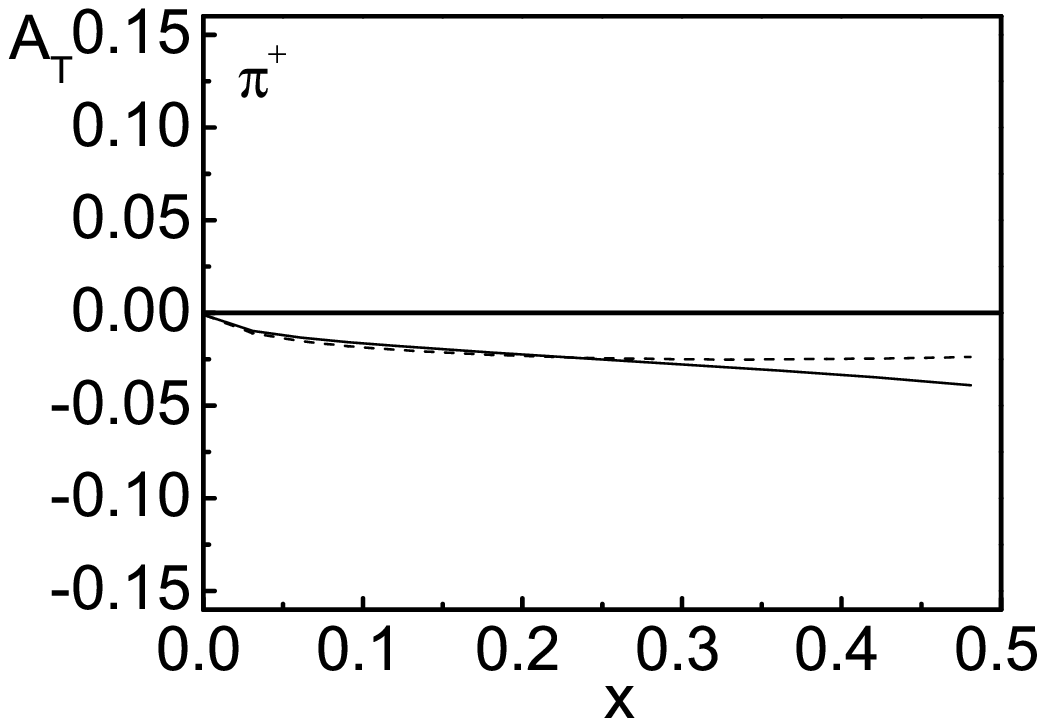}
\includegraphics[width=8cm]{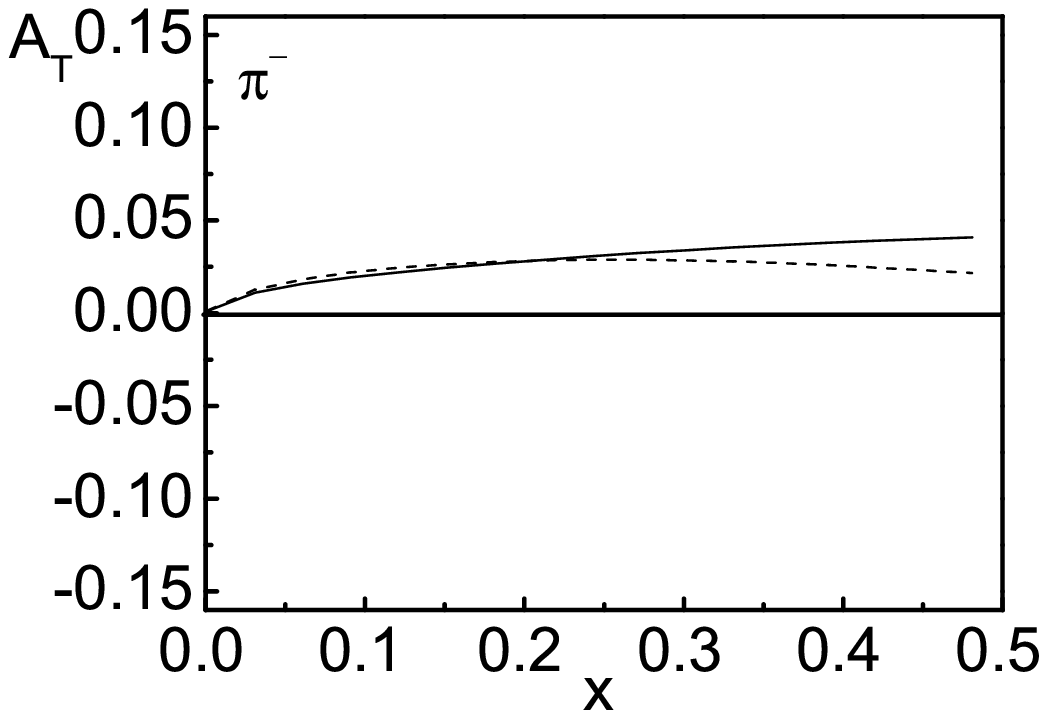}\nonumber\\
\includegraphics[width=8cm]{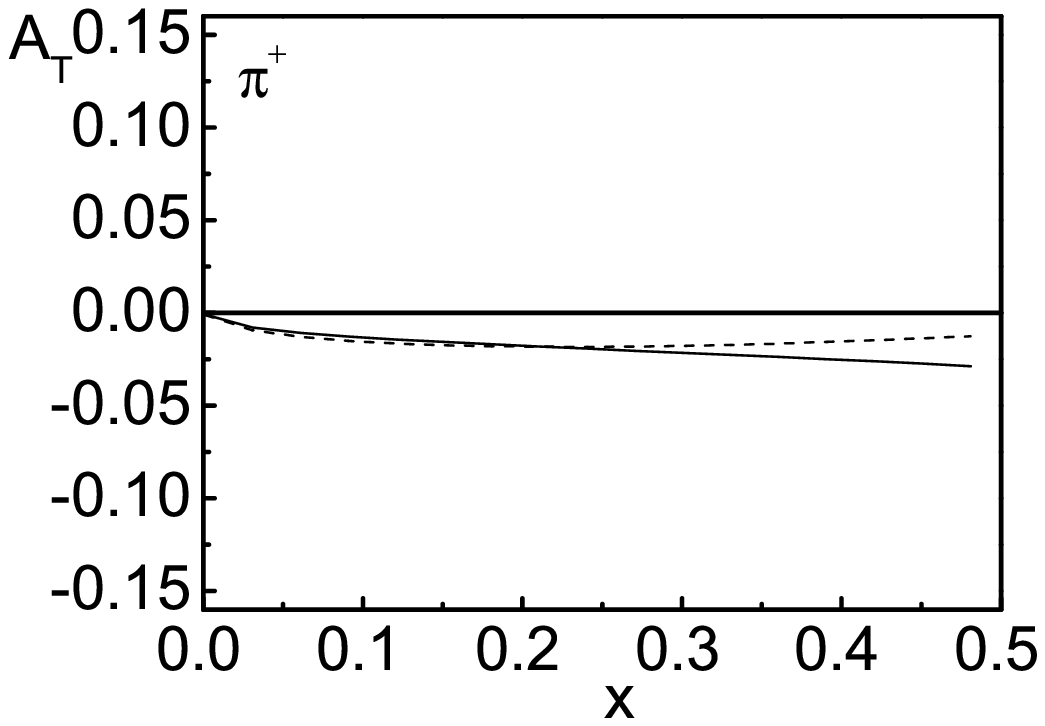}
\includegraphics[width=8cm]{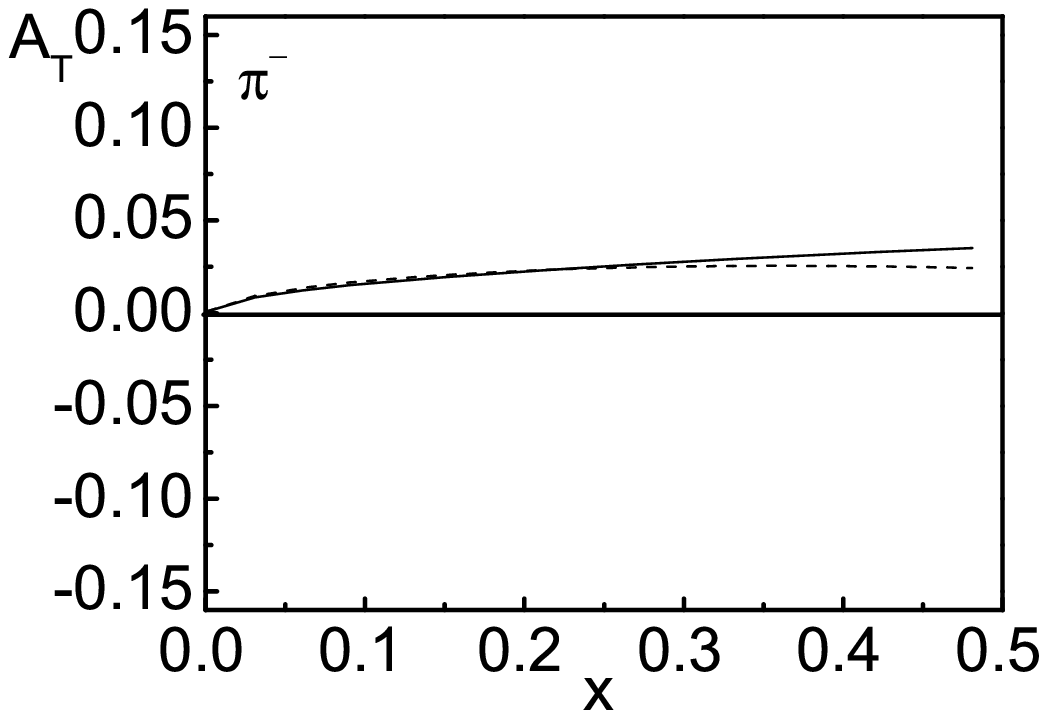}

\caption{Same as Fig. {\ref{fig3}}, but for the $neutron$ target.}
\label{fig4}
\end{figure}

\begin{figure}
\center
\includegraphics[width=8cm]{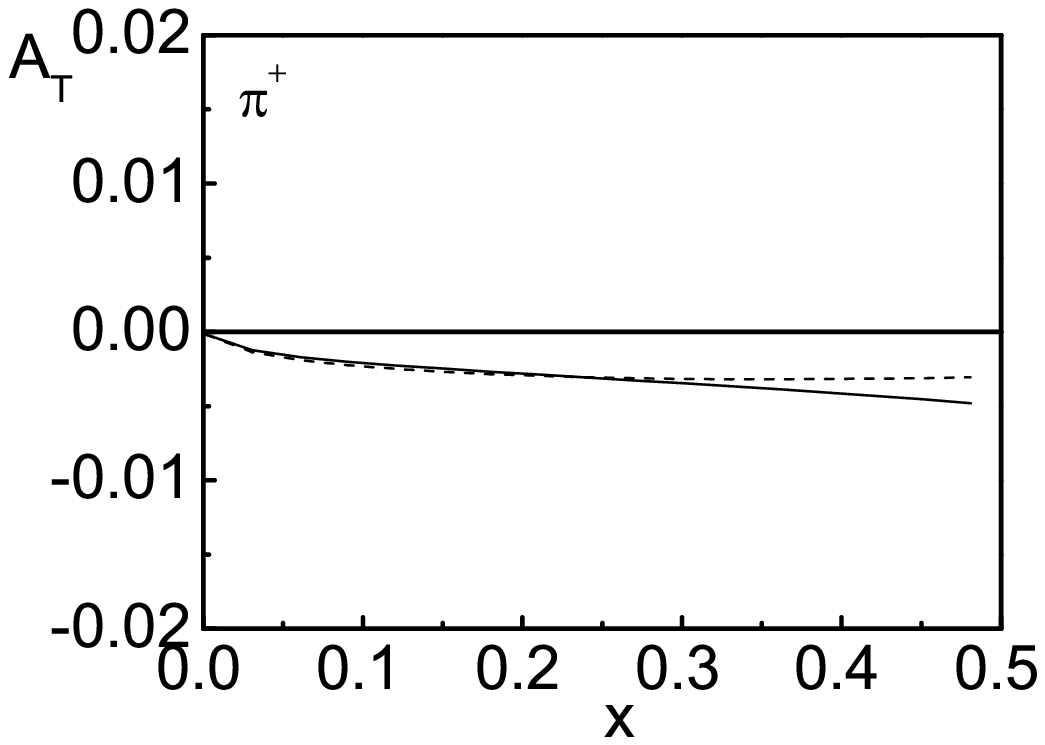}
\includegraphics[width=8cm]{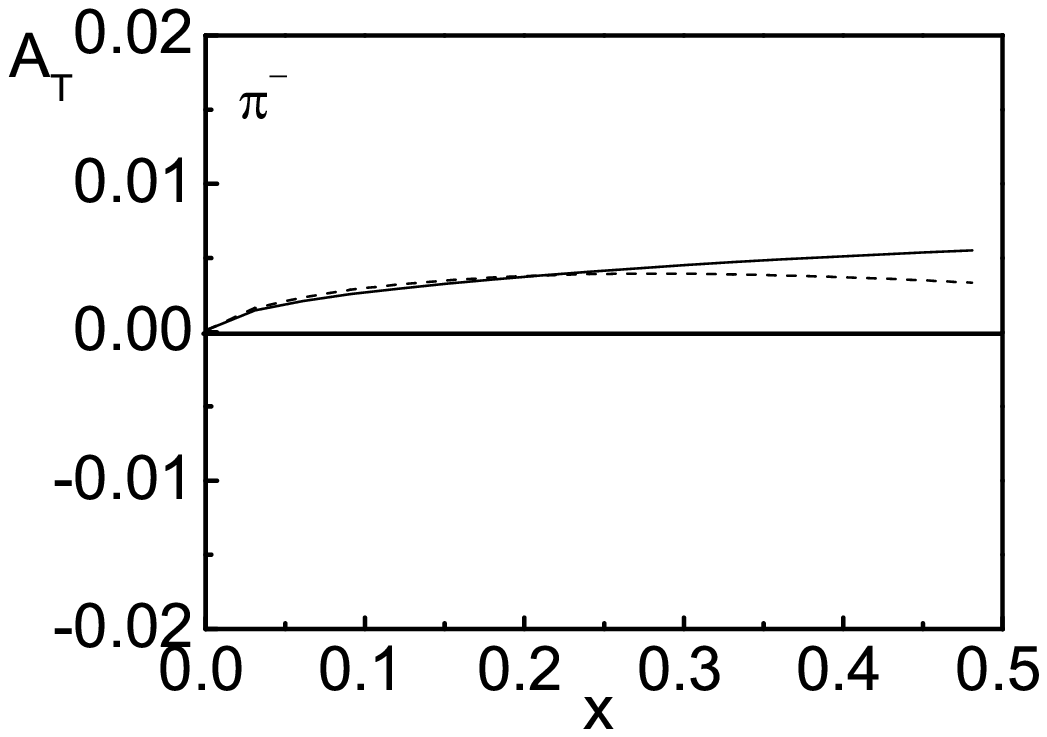}\nonumber\\
\includegraphics[width=8cm]{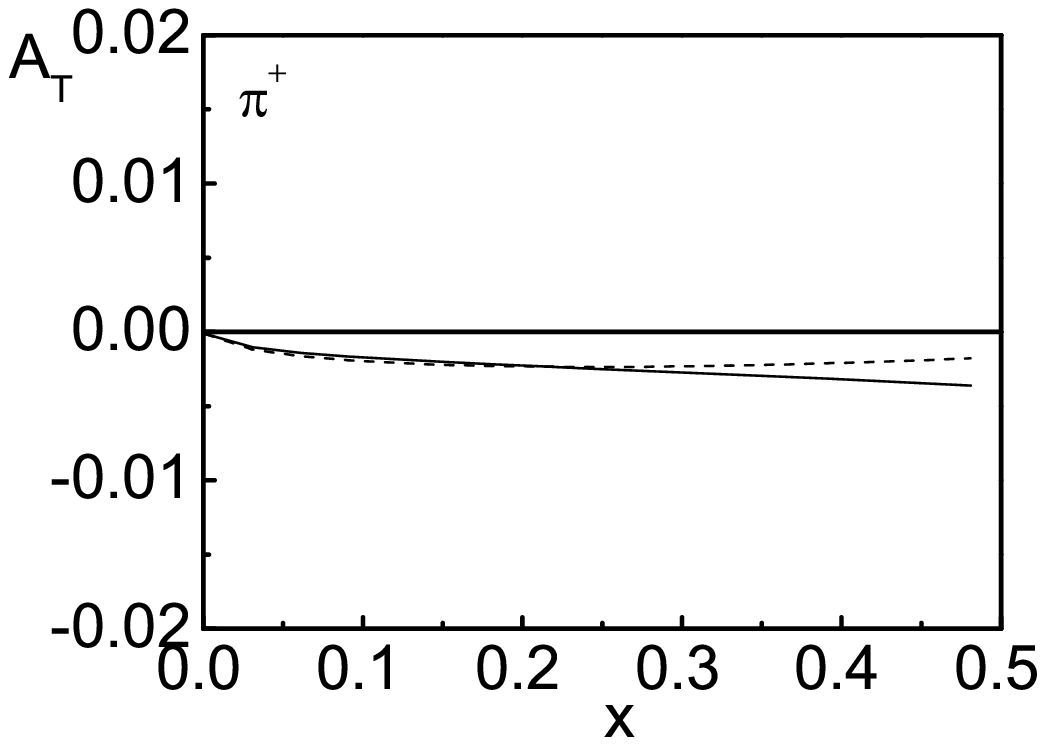}
\includegraphics[width=8cm]{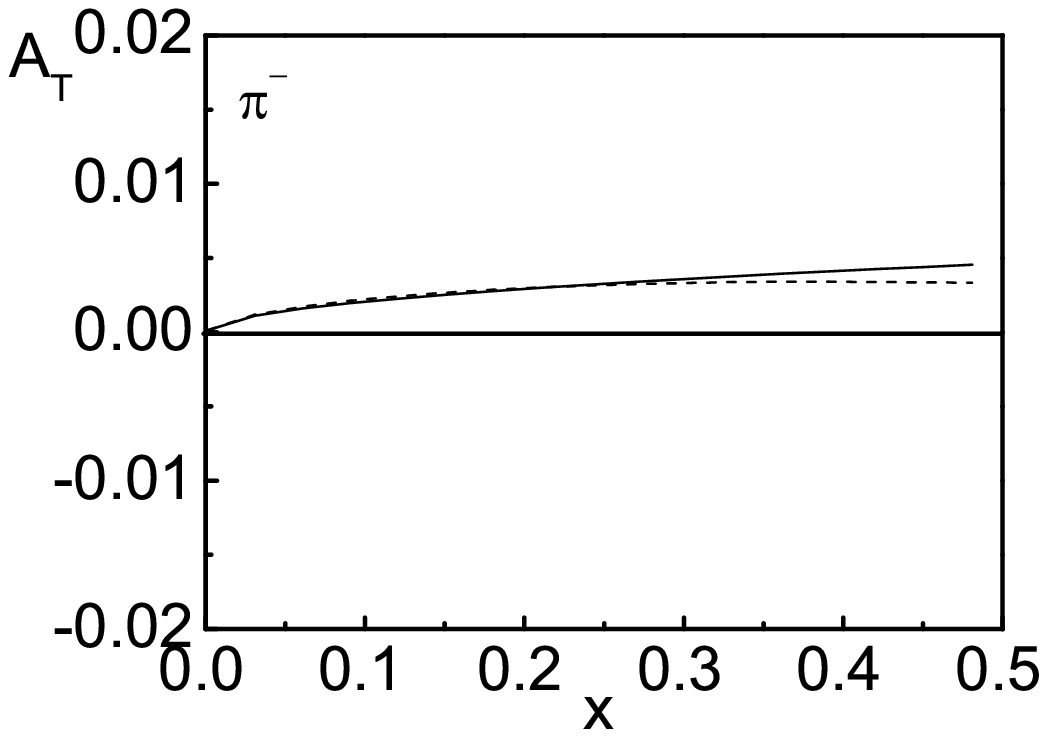}
\caption{Same as Fig. {\ref{fig3}}, but for the $^3$He target.}
\label{fig5}
\end{figure}

Fig.~\ref{fig3} shows the prediction of the asymmetry ($A_T$) on
proton target. Compared to Fig.~\ref{fig2}, we find that the
predictions are almost the same. Fig.~\ref{fig4} shows the neutron
case, where $d$ quark dominates, so the neutron data will be more
sensitive to the d quark in comparison with the hydrogen target.
Here $d$ to $\pi^-$ is favored and $d$ to $\pi^+$ is unfavored.
Again we get the result that the two processes both have the sizable
effect, although the asymmetry in $\pi^+$ production seems not
larger than that in the $\pi^-$ production as we argued in the above
section due to the different electric charges between $u$ and $d$
quarks. Fig.~\ref{fig5} is for the $^3$He target, and we expect a
similar prediction as the neutron case shown in Fig.~\ref{fig4}, for
$^3$He can be considered as a nearly free polarized neutron. But the
magnitude of the asymmetry is much smaller than the proton result as
the HERMES experiments showed, because some small coefficients
should be multiplied according to Eq.~(\ref{He}). This is the
difficulty measuring the structure of neutron for there is no such a
free neutron as a proton. We expect precise experiments that will be
done at JLab can give us some information on neutron.

 From Fig.~\ref{fig3} to Fig.~\ref{fig5}, we find that the
quark-diquark model and the pQCD based analysis still give similar
predictions. Also the two sets of parametrization predict similarly.
So the pion case is insensitive to the models, thus can give
constraints on both the Collins functions and transversity
distributions that are little known yet, rather than give direct
information on transversity distributions.

\begin{figure}
\center
\includegraphics[width=8cm]{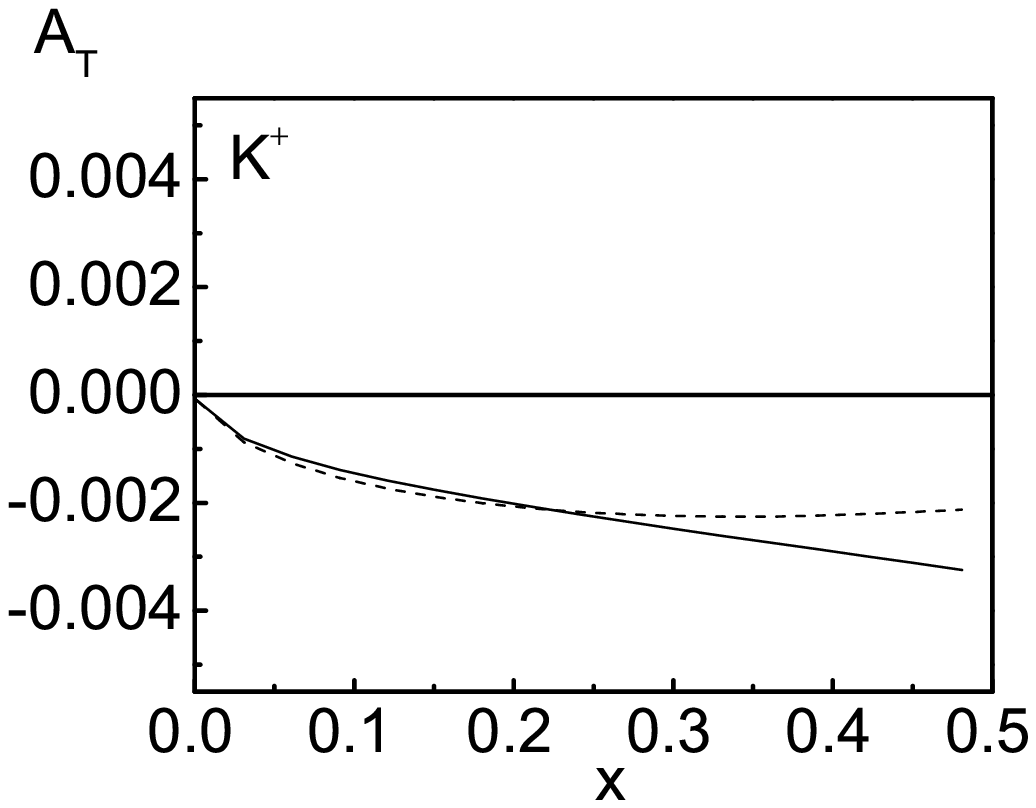}
\includegraphics[width=8cm]{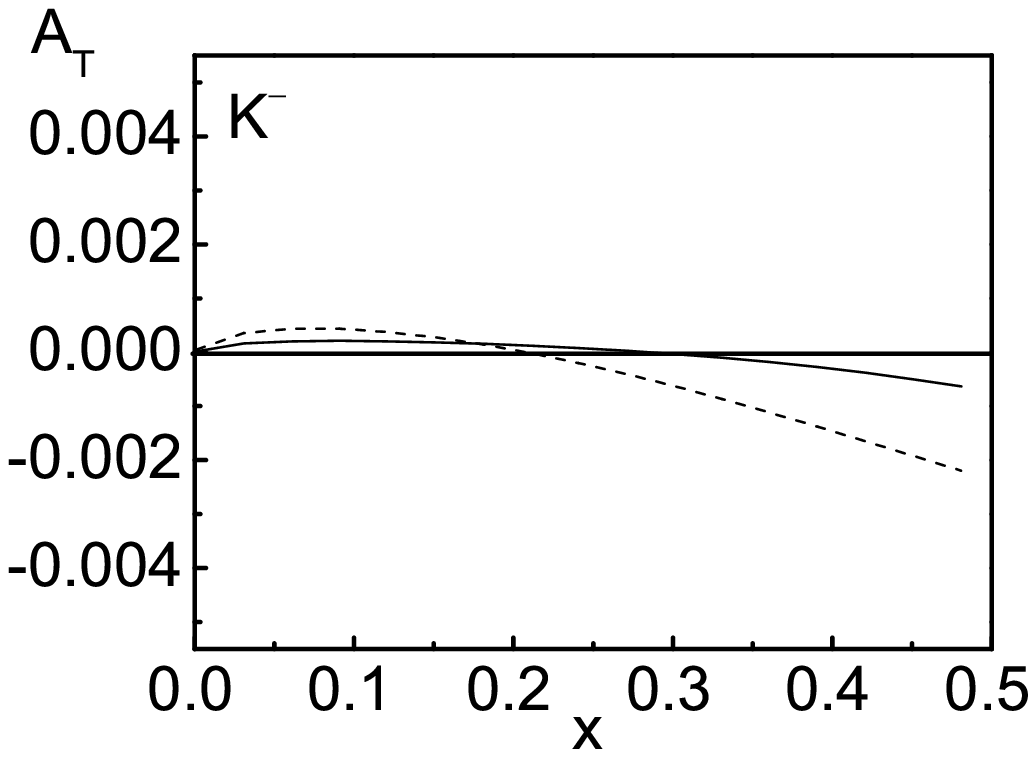}\nonumber\\
\includegraphics[width=8cm]{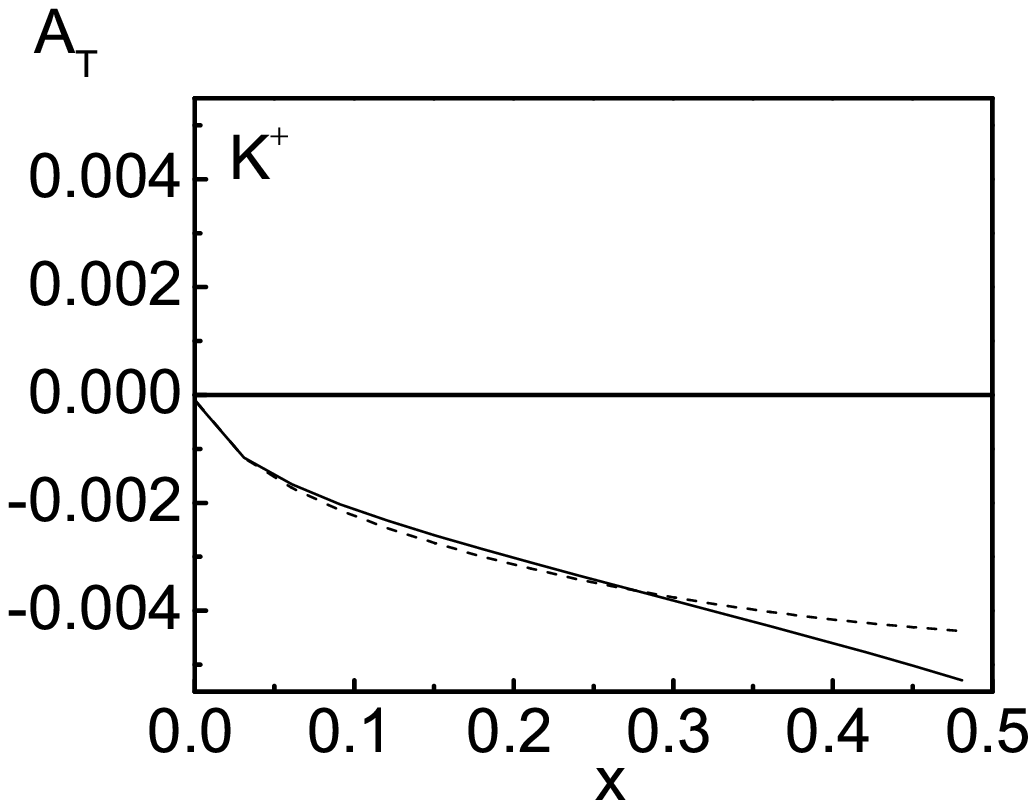}
\includegraphics[width=8cm]{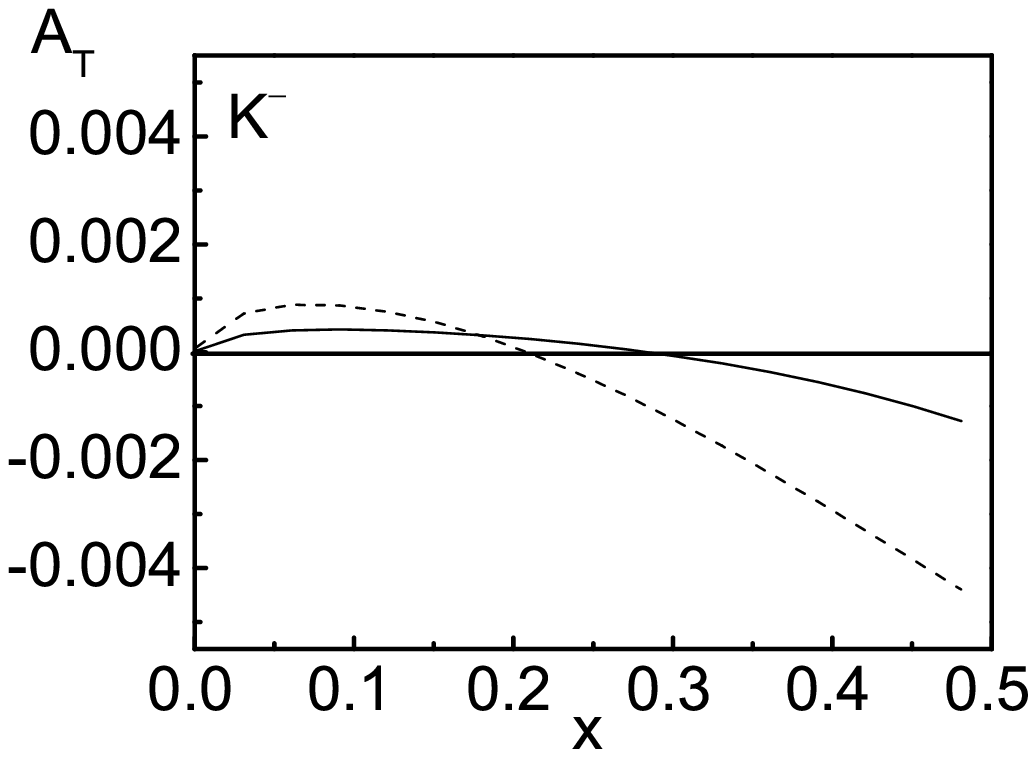}
\caption{Same as Fig. {\ref{fig5}}, but for the $K^+$ and $K^-$
productions.} \label{fig6}
\end{figure}

Next we present the predictions on $K^{\pm}$ productions as
Fig.~\ref{fig6} shows. $K^+$ comprises a $u$ and an $\bar{s}$ quark,
thus the main contribution for the $K^+$ production comes from the
valence $u$ quark with the favored process and the valence $d$ quark
with the unfavored process. This is quite similar to the case of
$\pi^+$ production. But for $K^-(\bar{u}s)$, the result is quite
different from the others. $K^-$ comprises a $\bar{u}$ quark and an
$s$ quark, so the favored process do not contribute for the case,
because the transversity is mainly the behavior of valence quarks.
Consequently, the asymmetry of $K^-$ production only comes from the
unfavored process. So the kaon production can give us information
about the unfavored Collins function, and we are looking forward to
the measurement.

The predictions on kaon productions are shown in Fig.~\ref{fig6}.
This time, we can see clearly that the quark-diquark model and the
pQCD based analysis still give similar results on $K^+$, but the two
different approaches of Collins functions seem to have differences.
Thus we expect the experiments to give constraints on the Collins
functions through $K^+$ production. For the $K^-$ case, not only the
two different approaches of Collins functions, but also the two
models of distributions predict differently. So it is concluded that
we can get information on Collins functions through $K^+$
production, and then distinguish the two models of distributions
through $K^-$ production. However we should notice that the
asymmetries for kaon productions are very small, and the differences
between the models are also small, approximately the same order as
the asymmetries. Since the measurements at JLab cannot reach such
precision, we have to admit that distinguishing models through
experiments is still difficult, and high precision experiments are
needed to clarify the details.

The results of JLab on a polarized $^3$He target can give
complementary results to HERMES hydrogen measurement. Compared with
COMPASS deuterium measurement which is an indirectly measurement of
neutron, JLab data will have a unique advantage because of the
higher Bjorken \textit{x}, since the transversity property is mainly
a valence behavior. In addition, we expect that JLab results will
give more detailed information on fragmentation functions.

\section{Summary}

Using quark-diqurk model and pQCD based analysis for distributions,
and new sets of parametrization of Collins functions, we reanalyzed
the HERMES experiment and found that with the new parametrization of
the Collins functions advanced by~\cite{fe:2005}, the predictions
are consistent with the new released HERMES data~\cite{HERMES1},
which implies that the ``unfavored'' Collins functions may play an
important role.

Furthermore, we calculated the azimuthal spin asymmetries of pion
and kaon productions in semi-inclusive deep-inelastic scattering of
an unpolarized charged lepton beam on a transversely polarized
$^3$He target in the JLab kinematics region. Due to the lack of
independent measurement of Collins functions, it is difficult to
obtain transversity distribution functions directly from the
measurement on pion productions. The comparison of the data with the
prediction can give constraints on both the Collins functions and
transversity distributions, but not a direct measurement of the
transversity distributions.

Using different models for distribution functions and Collins
functions , we found that the $K^+$ production is sensitive to
different sets of Collins functions but insensitive to different
models of distributions and the $K^-$ production is sensitive to
both different approaches of distributions and Collins functions. So
we suggest distinguishing the quark-diquark model and pQCD based
analysis through this process after the Collins functions being
constrained by the $K^+$ production. But due to the small magnitude
of the asymmetry, the measurement through this suggestion can only
be achieved by experiments with much higher precision compared to
the coming JLab experiment. The $\pi^{\pm}$ productions are
sensitive to neither the different approaches of distributions and
Collins functions, so they might not give much exciting results, but
these processes can be useful to give constraints on the results.
Additionally, we point here that the asymmetry of $K^-$ production
is contributed by pure unfavored processes, thus is an ideal process
to study the unfavored Collins functions.

%%%%%%%%%%%%%%%%%%%%%%%%%% Fig. 1  proton target%%%%%%%%%%%%%%%%
%%%%%%%%%%%%%%%%%%%%%%%%%%%%%%%%%%%%%%%%%%%%%%%%%%%%%%%%%%%%%%%

%%%%%%%%%%%%%%%%%%%%%%%%%% Fig. 2  proton target%%%%%%%%%%%%%%%%
%%%%%%%%%%%%%%%%%%%%%%%%%%%%%%%%%%%%%%%%%%%%%%%%%%%%%%%%%%%%%%%

%%%%%%%%%%%%%%%%%%%%%%%%%% Fig. 3  proton target%%%%%%%%%%%%%%%%
%%%%%%%%%%%%%%%%%%%%%%%%%%%%%%%%%%%%%%%%%%%%%%%%%%%%%%%%%%%%%%%

%%%%%%%%%%%%%%%%%%%%%%%%%% Fig. 4  proton target%%%%%%%%%%%%%%%%
%%%%%%%%%%%%%%%%%%%%%%%%%%%%%%%%%%%%%%%%%%%%%%%%%%%%%%%%%%%%%%%

%%%%%%%%%%%%%%%%%%%%%%%%%% Fig. 5  proton target%%%%%%%%%%%%%%%%
%%%%%%%%%%%%%%%%%%%%%%%%%%%%%%%%%%%%%%%%%%%%%%%%%%%%%%%%%%%%%%%

%%%%%%%%%%%%%%%%%%%%%%%%%% Fig. 6  proton target%%%%%%%%%%%%%%%%
%%%%%%%%%%%%%%%%%%%%%%%%%%%%%%%%%%%%%%%%%%%%%%%%%%%%%%%%%%%%%%%

\section*{Ackonwledgments}
We are grateful to Xiaodong Jiang, Haiyan Gao, and Xin Qian for
useful discussions. This work is partially supported by National
Natural Science Foundation of China (Nos.~10421503, 10575003,
10528510), by the Key Grant Project of Chinese Ministry of Education
(No.~305001), by the Research Fund for the Doctoral Program of
Higher Education (China).

%\newpage

\end{document}